\documentclass[aps,nofootinbib,showpacs,twocolumn,floatfix]{revtex4}
\usepackage{graphicx}
\usepackage{epsf,amsfonts,amssymb,amsbsy}
\usepackage[mathscr]{eucal}
\begin{document}
\title{Quantum spectrum of black holes }
\author{V.V.Kiselev}
\email{kiselev@th1.ihep.su}
\affiliation{Russian State Research Center ``Institute for High
Energy Physics'', 
Pobeda 1, Protvino, Moscow Region, 142281, Russia\\ Fax:
+7-0967-744937}
\pacs{04.70.Dy}
\begin{abstract}
The black hole as the thermodynamical system in equilibrium
possesses the periodicity of motion in imaginary time, that allows
us to formulate the quasi-classical rule of quantization. The rule
yields the equidistant spectrum for the entropy of non-rotating
black holes as well as for the appropriately scaled entropy in the
case of rotation. We clarify and discuss a role of quasi-normal
modes.
\end{abstract}
\maketitle


\section{Introduction}

Recently, we have used the technique of quantum thermal geodesics
confined behind horizons of black hole \cite{K1,K2,K3} in order to
derive a quasi-classical spectrum of masses for rotating Kerr and
BTZ black holes \cite{K2', K-BTZ}.

For the Kerr black hole, we have found the following relation
between the orbital momentum $J$ and mass $M$ \cite{K2'}:
\begin{equation}\label{Kerr-J}
    J = \frac{2\sqrt{l}}{l+1}\,M_{{\textsc{\scriptsize Kerr}}}^2,
\end{equation}
where we put the gravitational constant equal to unit ($G=1$). The
parameter $l$ takes the values of
\begin{equation}\label{Kerr-l}
    l=\{1,\, \textstyle{\frac{3}{2}},\, 2,\, 3,\,\infty\},
\end{equation}
representing the ratio of external horizon area to the internal
one
$$
l=\frac{{\cal A}_+}{{\cal A}_-},
$$
as follows from the consistency of mapping the analytically
continued space for radial geodesics completely confined behind
the horizons. Therefore, the Kerr black hole with respect to
rotation and radial motion has got two quantum numbers: the first
is the integer or, generically, half integer momentum $J$, while
the second is the `loop number' $l$. It is important to emphasize
that the quantum spectrum of Kerr black hole possesses the
loop-duality: the spectrum is invariant under the action of
duality transform
$$
l \leftrightarrow \frac{1}{l}.
$$
The extremal black hole corresponds to $l=1$. The limit of
$l\to\infty$ gives the Schwarzschild black hole, so that $J\to
2M^2/\sqrt{l}\to 0$, which indicates a breaking down such the
quantization method in the case of Schwarzschild black hole.

The same note concerns the BTZ black hole \cite{BTZ}, in which
case we have found the following spectrum \cite{K-BTZ}:
\begin{equation}\label{BTZ-J}
    J=\frac{2 k}{k^2+1}\, M_{{\textsc{\scriptsize btz}}}\ell,
\end{equation}
where $\ell$ is the curvature radius of AdS$_3$ space-time, while
$k$ is the loop number for the BTZ black hole. The loop-duality
$$
k\leftrightarrow \frac{1}{k}
$$
remains the spectrum invariant. Again, the non-rotating limit of
$k\to \infty$ misses the quantization in the form of
(\ref{BTZ-J}).

Therefore, for non-rotating black holes we need a consistent
quantization procedure supplemental to the case of $J\neq 0$. In
section II we use the quasi-classical method for the periodic
motion in purely imaginary time, which corresponds to the
thermodynamical ensemble, that is the case of geodesics confined
behind the horizons. The offered approach is applied to the
Schwarzschild black hole in section III and to the BTZ black hole
in section IV. We compare the procedure of quantization at $J=0$
with that of $J\neq 0$ and clarify the difference. We make notes
on the connection of quantization with quasi-normal modes (see
reviews in \cite{QNM-rev}). Several remarks are devoted to the
comparison with another quantization procedure developed by
\cite{BarKust}. Our results are summarized in Conclusion.

\section{Quasi-classical method and thermodynamical ensemble}

Let us start with a system possessing the only dynamical degree of
freedom, the generalized coordinate $q$, moving periodically.
Then, the quasi-classical quantization rule is the following:
\begin{equation}\label{quasi-1}
    \oint p\,{\rm d}q =2\pi\hbar\,n,\qquad n\in \mathbb N,
\end{equation}
where $p$ is the momentum canonically conjugated to $q$, while
$n\gg 1$ is the quantum number. In (\ref{quasi-1}) we have
neglected a possible shift of $n$ due to reflections at
turn-points. The energetic density of levels ${\rm d}n/{\rm d}E$
can be easily derived by differentiating (\ref{quasi-1}) with
respect to energy $E$, so that
\begin{equation}\label{quasi-2}
    \frac{{\rm d}n}{{\rm
    d}E}=\frac{1}{2\pi\hbar}\oint\frac{\partial p}{\partial E}\,{\rm
    d}q.
\end{equation}
The Hamilton equations give
\begin{equation}\label{quasi-3}
    \frac{\partial E}{\partial p}=\dot q,
\end{equation}
where $\dot q={\rm d}q/{\rm d}t$ is the velocity of motion.
Therefore,
\begin{equation}\label{quasi-4}
    \frac{{\rm d}n}{{\rm d}E} =\frac{\tau(E)}{2\pi\hbar},
    \qquad \tau(E)=\oint{\rm d}t,
\end{equation}
where $\tau$ is the period of motion depending on the energy.
Introducing the phase frequency
$$
\omega(E)=\frac{2\pi}{\tau(E)},
$$
we get
\begin{equation}\label{quasi-5}
    \frac{{\rm d}n}{{\rm d}E}=\frac{1}{\hbar\omega(E)}.
\end{equation}
Thus, we have reminded the ordinary result on the spacing between
the levels of energy: $\Delta E =\hbar\omega(E)\,\Delta n$.

Summing up the number of levels in a given interval of energy we
get
\begin{equation}\label{quasi-6}
    n\hbar= \int \frac{{\rm d}E}{\omega(E)},
\end{equation}
or equivalently
\begin{equation}\label{quasi-7}
    2\pi\hbar\,n=\int \tau(E)\,{\rm d}E.
\end{equation}

Next, in a thermodynamical equilibrium, a system is moving
periodically in purely imaginary time, so that the period is fixed
by the inverse temperature $\beta=1/T$. Therefore, we have got the
substitution
\begin{equation}\label{quasi-t1}
    \tau(E)\mapsto-{\rm i}\hbar\,\beta({\cal E}),\qquad
    E\mapsto {\rm i}\,{\cal E},
\end{equation}
with $\cal E$ denoting the Euclidean energy. So, we derive
\begin{equation}\label{quasi-t2}
    \int \beta({\cal E})\,{\rm d}{\cal E}=2\pi\,n,\qquad n\in
    \mathbb N.
\end{equation}

In the framework of quantum thermal geodesics confined behind the
horizon, the black hole is represented by definite microstates,
namely, the system of particles on geodesics in analytically
continued space defined behind the horizon. Each particle is
ascribed to a winding number $n_W$ determining the number of
cycles per period, i.e. $\beta_{n_W}=\beta/n_W$ \cite{K1,K2,K3}.
Summing up contributions by microstates is equal to summing over
the particles and cycles:
\begin{equation}\label{quasi-bh1}
    \sum\limits_{micro.}\int \beta({\cal E})\,{\rm d}{\cal
    E}\Big|_{micro.}\equiv
    \sum\limits_{part.}\sum\limits_{cycle}^{n_W}\int
    \frac{\beta({\cal E})}{n_W}\,{\rm d}{\cal E}.
\end{equation}
Since
$$
\sum\limits_{cycle=1}^{n_W}\frac{\beta}{n_W}=\beta,
$$
we get the same overall period, while the summing over particles
gives the total energy of the black hole, i.e. its mass \cite{K3}.
Therefore, for the non-rotating black hole with the single
external characteristics, the mass $M$, the quantization rule of
(\ref{quasi-t2}) takes the form
\begin{equation}\label{quasi-bh2}
    \int\beta(M)\,{\rm d}M=2\pi\,n,\qquad n\in \mathbb N.
\end{equation}
Using the thermodynamical relation
$$
{\rm d}M= T\,{\rm d}\mathcal S,
$$
where $\mathcal S$ is the entropy, we arrive to
\begin{equation}\label{quasi-bh3}
    \mathcal S=2\pi\,n.
\end{equation}
Thus, the quasi-classical quantization of non-rotating black hole
results in the equidistant quantization of its entropy\footnote{We
do not consider charged black holes, too, since the derivation has
been based on the fact of single dynamical quantity of black hole,
its mass.}, which is in agreement with the argumentation by
J.Bekenstein in his pioneering paper on the quantum spectrum of
black hole area \cite{Bekenstein} as well as with further
developments in \cite{BM}.

\section{Kerr black hole: $J=0$ and $J\neq 0$}

\subsection{Schwarzschild black hole}

Non-rotating black hole satisfies the condition of single
dynamical variable, the black hole mass. Therefore, we
straightforwardly get the quasi-classical spectrum
\begin{equation}\label{Schwarz-1}
    \mathcal S_n =4\pi M_n^2=2\pi\,n\quad\Rightarrow\quad
    M_n^2=\frac{n}{2}.
\end{equation}
This result coincides with the quasi-classical limit of spectrum
obtained in \cite{BarKust} in the framework of quantizing the
effective dynamical system of black hole in terms of its global
external characteristics. Then, after canonical transformation one
gets the quantum system equivalent to harmonic oscillator. This
transform uses the canonically conjugated pair of black hole mass
$M$ and periodic angle-like variable $P$ as conjectured by authors
of \cite{BarKust}. However, we can point out a problem related
with a constructing of Hermitian phase operator conjugated to the
occupation number of oscillator in quantum mechanics as reviewed
in \cite{CarrNieto}. Nevertheless, the problem is irrelevant in
the quasi-classical approximation, but it is important while exact
quantization. In addition, the oscillator-like spectrum in
\cite{BarKust} yields the entropy, not the total energy, that
could change the situation. Thus, the result of this section is in
agreement with that of obtained by the method of \cite{BarKust}.

Similar ideas were used by H.Kastrup in \cite{Kastrup}. However,
the obtained spectrum $M_n^2=n/4$ includes the additional one
half, which reflects a systematic miss caused by the heuristic
correspondence of energy multiplied by the period in imaginary
time, with the quantized adiabatic invariant. So, we have improved
this guess by more strict argumentations. Ref. \cite{Kastrup}
contains a discussion on the black hole entropy and
thermodynamics, too.

\subsection{Quasi-normal modes and quantum spectrum}

Remarkably, one could explore independent determination of phase
frequency $\omega(E)$ in order to use the quantization rule in the
form of (\ref{quasi-6}), which is a formal expression of Bohr's
correspondence principle: classical frequencies reproduce
increments of energy between the levels at high quantum numbers,
i.e. $\Delta E =\hbar\omega(E)\,\Delta n$. Such classical
frequencies correspond to quasi-normal modes \cite{QNM-rev}. Those
modes have got both real and imaginary terms (see original
evaluations in \cite{QNM-original}, recent analytical results were
obtained in \cite{Schiappa,Schiappa2}). In
\cite{QNM-adiabat,Setare} authors use the real parts in order to
quantize the black hole spectrum. However, as we have argued in
section II, the thermodynamical system is inherently periodic with
imaginary time. This fact is exactly reproduced by tower of the
imaginary parts in quasi-normal modes. So, we insist that
procedure based on the quasi-normal modes is consistent with the
quantization performed above, if only one uses the imaginary part
of classical frequencies, which are universal for classical fields
with various spins, while the substitution of real parts of
quasi-normal modes in quantization rule (\ref{quasi-6}) seems to
be misleading. Nevertheless, the real parts of quasi-normal
frequencies could contain some other physical information.

This fact could be especially important in connection with
attempts to use Bohr's correspondence principle in the framework
of Loop Quantum Gravity (see recent reviews in \cite{LQG} and
references therein) in order to fix both Immirzi parameter and
quantum spacing of black hole horizon area \cite{Hod,Dreyer}. In
that case one exploits the real parts of quasi-normal modes to
relate it with the area law and minimal value of spin in its
network. Unfortunately, to our opinion, again, the substitution of
real parts of quasi-normal frequencies in Bohr's correspondence
principle is misleading in the context of black hole
thermodynamics.

In addition, authors of \cite{Schiappa,Schiappa2} argue for the
real parts of quasi-normal modes cannot be straightforwardly
applied to other black holes except the simplest case of
Schwarzschild black holes in the context of Loop Quantum Gravity.
Such the argumentation invalidates some proposals in
\cite{Dreyer}. Moreover, in \cite{Schiappa2} one finds a
discussion, why real parts of asymptotic quasi-normal modes cannot
be used in semi-classical considerations of Loop Quantum Gravity,
at suggested in \cite{Hod,Dreyer}.

\subsection{Kerr black hole}

At $J\neq 0$ we have two dynamical variables, the mass $M$ and
orbital momentum $J$, so one has to take into account the
quantization of angular motion. Moreover, it is important to pay
attention to both horizons, the external and internal ones.
Indeed, the angular velocities of horizons are equal to
\begin{equation}\label{Kerr-1}
    \Omega_\pm=\frac{a}{r_\pm^2+a^2},
\end{equation}
where $a=J/M$, and $r_\pm$ are the radii of horizons. The
quantization of horizon-area ratio leads to strict relation
between the mass and orbital momentum shown in (\ref{Kerr-J}).
Then, $M$ and $J$ are not independent at fixed loop $l$ of
(\ref{Kerr-l}), that makes the single-variable quantization of
(\ref{quasi-6}), (\ref{quasi-7}) or (\ref{quasi-bh2}),
(\ref{quasi-bh3}) irrelevant. Under relation (\ref{Kerr-J}) we get
\begin{equation}\label{Kerr-2}
    \Omega_+=\frac{1}{2\sqrt{l}M},\qquad \Omega_-=l\,\Omega_+.
\end{equation}
The temperatures at horizons are given by
\begin{equation}\label{Kerr-3}
    \beta_+=8\pi M\,\frac{l}{l-1},\qquad
    \beta_-=\frac{\beta_+}{l}.
\end{equation}
So, the self-dual angle of rotation per thermodynamical period is
equal to
\begin{equation}\label{Kerr-4}
    \Delta\phi=\beta_+\Omega_+=\beta_-\Omega_-=4\pi\,\frac{\sqrt{l}}{l-1}.
\end{equation}
The corresponding winding numbers for the ground state at horizons
are given by
\begin{equation}\label{Kerr-5}
    n_W^+=\frac{2l}{l-1},\qquad
    n_W^-=\frac{n_W^+}{l}=\frac{2}{l-1}.
\end{equation}
Introduce a multiple period consistent for both horizons,
\begin{equation}\label{Kerr-6}
    \tau=\sqrt{\beta_+\beta_-}=\frac{\beta_+}{\sqrt{l}},
\end{equation}
which gives the following rotation angles
\begin{equation}\label{Kerr-7}
    \begin{array}{ccc}
      \Delta\phi_+ & =\tau \Omega_+ & \displaystyle=2\pi\,\frac{2}{l-1}=2\pi\,n_W^-,
      \\[4mm]
      \Delta\phi_- & =\tau \Omega_- & \displaystyle=2\pi\,\frac{2l}{l-1}=2\pi\,n_W^+.
    \end{array}
\end{equation}
Therefore, at both horizons the Kerr black hole makes rotations by
angles multiple to $2\pi$ per the specified time period. The
multiplication factors are identical to winding numbers. Thus, the
complete periodicity with account of rotation takes place at
$\tau=\beta_+/\sqrt{l}$. Note, that due to
$$
T\,{\rm d}\mathcal S={\rm d}M-\Omega_+{\rm d}J
$$
we can deduce
\begin{equation}\label{Kerr-S}
    \int\limits_0^M\tau(M){\rm d}M\,\left(1-\Omega_+\,\frac{{\rm d}J}{{\rm d}M}\right)
    =\frac{\mathcal S}{\sqrt{l}}=2\pi\,J,
\end{equation}
that provides the correct quantization of entropy $\mathcal S$ as
it was obtained in \cite{K3}. Thus, we should modify the
quantization rule of (\ref{quasi-bh3}) by
\begin{equation}\label{QR}
    \frac{\tau}{\beta}\,\mathcal S=2\pi\,n,
\end{equation}
valid in the case of rotation, though $n$ could be a subset of
integer numbers.

In the quasi-classical approach, the horizon area spectrum is
given by
\begin{equation}\label{Kerr-8}
    \mathcal A = 8\pi\,\sqrt{l}\,J.
\end{equation}
In the same limit, formula (\ref{Kerr-8}) reproduces the spectrum
obtained in \cite{BarKust}, if only one puts the loop $l=1$. To
our opinion, the reason for such the correspondence is
transparent: if one ignores the dynamics on inner horizon (as in
\cite{BarKust}), one gets the consistent quantization supposing a
coherent rotation of both horizons, i.e. putting $l=1$.

Finally, it is interesting to note, that combining the cases of
$J=0$ and $J\neq 0$ at $l=1$, one could ascribe the spectrum of
$M^2=n/2$ to points of `daughter trajectories' of main trajectory
$J=M^2$ in the plane of $\{M^2,J\}$.

\section{BTZ black hole: $J=0$ and $J\neq 0$}

At $J=0$ we use the quantization rule of (\ref{quasi-bh3}) to
deduce
\begin{equation}\label{BTZ-1}
    \mathcal S_n= 2\pi\ell\sqrt{\frac{M_n}{2G}}=2\pi\,n,\quad
    M_n=2G\,\frac{n^2}{\ell^2}.
\end{equation}
The horizon area spectrum $\mathcal A_0=4G\,\mathcal S_n$ is also
equidistant.

At $J\neq 0$, after taking into account the spectrum of
(\ref{BTZ-J}), we find that the horizons rotate with angle
velocities
\begin{equation}\label{BTZ-2}
    \Omega_+ =\frac{1}{\ell k},\qquad \Omega_-=k^2\Omega_+,
\end{equation}
while the corresponding temperatures are given by
\begin{equation}\label{BTZ-3}
    \beta_+=\pi \ell\,
    \frac{k}{k^2-1}\sqrt{\frac{k\ell}{GJ}},\qquad
    \beta_-=\frac{\beta_+}{k}.
\end{equation}
Two horizons consistently rotate by angles multiple to $2\pi$ at
the period of
\begin{equation}\label{BTZ-4}
    \tau=4\pi\ell\,\frac{k}{k-1},
\end{equation}
so that the angles are determined by the winding numbers of ground
state at the horizons,
\begin{equation}\label{BTZ-5}
    \begin{array}{ccc}
      \Delta\phi_+ & =\tau \Omega_+ & \displaystyle=2\pi\,\frac{2}{k-1}=2\pi\,n_W^-,
      \\[4mm]
      \Delta\phi_- & =\tau \Omega_- & \displaystyle=2\pi\,\frac{2k^2}{k-1}=2\pi\,k\,n_W^+.
    \end{array}
\end{equation}
The ratio
\begin{equation}\label{BTZ-6}
    \frac{\tau}{\beta_+}=4(k+1)\sqrt{\frac{GJ}{k\ell}}
\end{equation}
gives
\begin{equation}\label{BTZ-7}
    \frac{\tau}{\beta_+}\,\mathcal S=2\pi J\,(2k+2),
\end{equation}
which is consistent with (\ref{QR}). The obtained result disagrees
with consideration in \cite{Setare}.

\section{Conclusion}

In the present paper we have used the periodic motion of
thermodynamical ensemble in imaginary time in order to formulate
the quasi-classical quantization rule for single-variable
dynamical system of black hole, i.e. non-rotating black hole. The
rule has given the equidistant quantization of entropy. The
application of method to Schwarzschild and BTZ black holes has
been considered. We have emphasized the difference with the
treatment in terms of quasi-normal modes: to our opinion the use
of real parts of frequencies is misleading in the problem under
study, while the imaginary parts of quasi-normal modes reproduce
our result. This fact makes irrelevant the treatment of quantum
spectrum for black holes in terms of Loop Quantum Gravity as
suggested in \cite{Hod,Dreyer} as well as in the quasi-classical
framework of \cite{QNM-adiabat}. Nevertheless, the real parts of
quasi-normal frequencies could have another physical sense.

We have clarified the difference of single-variable approach with
the case of rotating black holes. So, one has to take into account
consistent multiple folding of rotation for both horizons. This
consistency has required to scale the full period of motion for
the black hole as a whole. This scaling has resulted in the
modified quantization rule, which guarantees the appropriate
equidistant quantization of scaled entropy.

The mentioned consistency adjusting the rotation of both horizons,
was not generically taken into account in approach of
\cite{BarKust}, which, therefore, is theoretically sound at $l\to
1$, only, in the quasi-classical limit, since quantization of
periodic phase has some principal problems \cite{CarrNieto}.

This work is partially supported by the grant of the president of
Russian Federation for scientific schools NSc-1303.2003.2, and the
Russian Foundation for Basic Research, grant 04-02-17530.

\newpage

\end{document}